\documentclass[parallelisme]{compas2019}
\usepackage{amsmath}
\usepackage[T1]{fontenc}
\usepackage{algorithm}
\usepackage{algorithmic}
\usepackage{cite}
\usepackage{tikz}
\usepackage{pgfplots}
\usepackage{subfig}
\usepackage{graphicx}

\newcommand\VLONGUE{}

\toappear{1}

\def\RevVersion{}
\ifdefined \RevVersion
  \usepackage{soul}
  \usepackage{color}
  \usepackage{marginnote}
  \newcommand{\added}[1]{{\color{blue} #1}}
  \newcommand{\deleted}[1]{\st{#1}}
  \setstcolor{blue}
  
\else 
  \ifdefined \OldVersion
    \newcommand{\added}[1]{}
    \newcommand{\deleted}[1]{#1}
    
  \else
    \newcommand{\added}[1]{#1}
    \newcommand{\deleted}[1]{}
    
  \fi
\fi

\begin{document}
    \title{A WCET-aware cache coloring technique for reducing interference in real-time systems}
    \author{Fabien Bouquillon$^{1,2}$, Cl\'ement Ballabriga$^1$, Giuseppe Lipari$^1$, Smail Niar$^2$}
    \address{$^1$ Univ. Lille, CNRS, CentraleLille,
      UMR 9189 - CRIStAL, Lille, France\\[.2cm]
      $^2$ Univ. Polytechnique Hauts-de-France, LAMIH/CNRS,  Valenciennes, France
    }
    \maketitle
    \ifdefined\VLONGUE\else\vspace{-1em}\fi
    \begin{abstract}
  The time predictability of a system is the condition to give safe and precise bounds
  on the worst-case execution time of real-time functionalities which are running on
  it. Commercial off-the-shelf(COTS) processors are increasingly used
  in embedded systems and contain shared cache memory. This component
  has a hard predictable behavior because its state depends on the
  execution history of the systems. To increase the predictability of
  COTS component we use cache coloring, a technique widely used to
  partition cache memory. Our main contribution is a WCET aware
  heuristic which partition task according to the needs of each
  task. Our experiments are made with CPLEX an ILP solver with random
  tasks set generated running on preemptive system scheduled with
  earliest deadline first(EDF).
  
\end{abstract}

    \ifdefined\VLONGUE\else
    \let\thefootnote\relax\footnotetext{
      Acknowledgment:  This project is supported by \guillemotleft La Fédération de Recherche  Transports Terrestres \& Mobilité (FR TTM 3733) du CNRS \guillemotright \\
      Extended version of this work is available at: https://arxiv.org with the name A WCET-aware cache coloring technique for reducing interference in real-time systems }
    \fi
    \vspace{-1em}
    \section{Introduction}
\label{sec:introduction}

Hard real-time systems are found in many different domains, like
avionics, automotive, health care services. In such systems, a
real-time task has to be executed within predefined timing
constraints, whose violation can lead to system failure. Thus, it is
important to compute the response time of every task to ensure
\emph{a-priori} that it always executes within its time window under
all conditions. Schedulability analysis algorithms provide upper
bounds on the response time of tasks, which depend upon several
parameters such as tasks' execution times and scheduling policy. In
turn, execution times depend on the hardware architecture and task's
code.

Commercial off-the-shelf (COTS) processors are increasingly used in
embedded systems for their low cost and high performance. Most of COTS
processors use cache memories to bridge the gap between processor
speed and main memory access speed. In particular, cache memories
improve performance by reducing the \emph{typical} execution time of a
task.

However, in real-time systems we need predictability, that is we need
to precisely estimate the \emph{Worst-Case} Execution Time (WCET) of a
task.
Cache memory state depends on execution history of the system and, its prediction is a challenge due to the increase of tasks running on the system which compete for the same cache memory area.
More sophisticated WCET analyses take
into account the state of the cache during the execution of the task
and provide a tighter WCET. However, these analyses typically assume
that every task executes alone in the system, without interference
from other tasks. If tasks are executed concurrently and preemptively
on the system, one task may preempt another task and evict cache
blocks, making the estimated WCET too optimistic. This type of
interference is called \emph{inter-task} interference, as opposed to
\emph{intra-task} interference due to a task evicting its own cache
blocks.

In the literature, many researchers have been interested in this problem
by accounting for the cost of preemption through the so-called
\emph{Cache-Related Preemption Delay}
\cite{mancuso2013real,lunniss2012optimising,altmeyer2012improved
}. Another problem arises in multi-core systems with shared cache: a
task executing on one processor may evict useful cache blocks for a
seconds task executing on a different processor. It is, therefore,
necessary to reduce, or eliminate altogether, the inter-task
interference caused by the cache conflicts on tasks' execution times.

The goal of this research is to use virtual memory and cache-coloring
techniques to reduce inter-task interference: we allocate the virtual
pages of a task to physical pages so to minimize conflicts between
tasks on set-associative caches. Since cache memory is limited, by doing, 
so we might increase intra-task conflicts: two pages from the same
task may be allocated to two physical pages that correspond to the
same position in the cache, thus increasing the task's WCET. We, 
therefore, propose a methodology to explore the space of possible
cache-coloring configurations so to reduce conflicts while maintaining
the respect of the timing constraints. We can represent this problem
as a variant of the multiple-choice knapsack problem where the colors
are the knapsack and the pages the object, but in this variant, the
values of the object depend also on the presence of the other objects
in the same knapsack. Since the problem's complexity is very large, we
propose a combination of Integer-Linear-Programming techniques and
heuristics to partition the cache taking into consideration the WCET of each task.


    \section{Related Works}
\label{sec:related-works}

Predictability of cache memory in real-time systems has been widely
explored, especially for the CRPD
\cite{mancuso2013real,lunniss2012optimising,altmeyer2012improved }.

Luniss et al.\cite{lunniss2012optimising} used simulated annealing to
find a code layout in the memory that minimizes the CRPD. However,
tasks are not isolated on cache memory, so inter-task and inter-core
interference are still present. They used the linker to configure the
code layout.

Mancuso et al.~\cite{mancuso2013real} propose a complete framework
which defines, isolates and locks most important memory areas in
memory cache. These techniques are based on cache coloring partitioning
and cache locking, its purpose is to reduce conflicts and enhance
predictability but the cache is not optimally used because only the
most important memories area are in the cache and to access other areas
require costly RAM access. In our work, we use their techniques in
our heuristics for the pages coloring, but instead of giving all the
partitions to the most important data, we reserve a partition to the
other data.

Kim et al.~\cite{kim2013coordinated} propose a practical OS-level
cache management scheme using page coloring. They work on partitioned
fixed priority preemptive scheduling system where they partition cache
memory between cores with page coloring.
In their works tasks may share the same cache area, thus intra-core
interference is still present.

Ward et al.~\cite{ward2013outstanding} 
consider colors as shared resources protected by critical sections,
thus
priority inversion may occur during execution. 
To reduce this problem they propose to slice tasks' periods, but their
method may force the preempted task to reload its data (the set of
data pages that a task may access in one job).


\section{System model}
\label{sec:system-model}

In this section, we first present the task model, and then the model
of the hardware architecture.

We consider a system of $N$ real-time sporadic tasks
$\mathcal{T} = \{\tau_1, \cdots, \tau_N\}$. A task $\tau_i$ is an
infinite succession of jobs $J_{i,k}(a_{i,k}, c_{i,k}, d_{i,k})$, each
one characterized by an arrival time $a_{i,k}$, a computation time
$c_{i,k}$ and an absolute deadline $d_{i,k}$.  A job $J_{i,k}$ must be
executed in the interval of time $[a_{i,k}, d_{i,k}]$, if it misses
its deadline then a critical failure occurs.

A sporadic task $\tau_i$ can be represented by the tuple
$(C_i, D_i, T_i, P_i)$, where $C_i$ represents the worst case
execution time (WCET) of the task $\tau_i$
($C_i = \underset{\forall k, k \geq 1}{max}\{c_{i,k}\}$), $T_i$
represents the minimum time between two consecutive arrivals
($T_i \geq min_k \{a_{i,k + 1} - a_{i,k} \}$), $D_i$ is the relative
deadline ($\forall k, \; d_{i,k} = a_{i,k} + D_i$), and $P_i$ is the
number of distinct virtual pages used by the task.

We consider a set of sporadic tasks with
\ifdefined\VLONGUE
implicit deadline ($D_i = T_i$) or
\fi
constrained deadlines ($D_i \leq T_i$), scheduled with the preemptive
Earliest Deadline First (EDF) scheduler on a single processor. This
work can be easily extended to partitioned scheduling on a multi-core
system with shared caches. We also assume that tasks are independent,
that is they do not share any memory page. We will discuss later how
to remove such assumption.

We consider a set-associative cache and denote as $S^{instructions}$ the
number of distinct pages that fits into the cache. In this
paper we only focus on the \emph{instruction} cache: extension to data
cache is the subject of future work. We denote as $N^{way}$ the number of cache way.

The \emph{color} of the j-th virtual page $p_{i,j}$ of task $\tau_i$,
denoted as $\kappa_{i,j}$, is an index between $[0, \frac{ S^{instructions} }{N^{way}} - (N - 1)]$
that denotes the position in the cache where the page will be
loaded.
Therefore, we search a method for allocating virtual pages
to physical pages so that any two different tasks share the minimum
possible number of colors (ideally zero).

\begin{figure}
\begin{center}
  \subfloat[Cache colloring]{
\begin{tikzpicture}[thick,scale=0.25, every node/.style={transform shape}]

  \draw (1.5, 8.5) node {\Large TASK 1 VTABLE};
  \foreach \x in {0,...,2} {
    \draw (0, 5 + \x * 1 ) rectangle(3, 6 + \x * 1);
   \pgfmathtruncatemacro\addr{2 - \x}
  \draw (-0.5, 5.5 + \x * 1) node {\addr};
}
  
\draw (1.5, 2.5) node {\Large TASK 2 VTABLE};
\foreach \x in {0,...,1} {
  \draw (0, 0 + \x * 1 ) rectangle(3, 1 + \x * 1);
  \pgfmathtruncatemacro\addr{1 - \x}
  \draw (-0.5, 0.5 + \x * 1) node {\addr};
}
  
\draw (8,3.5) node {\Large MEMORY};
  \foreach \x in {0,...,7}
  \pgfmathtruncatemacro\myresult{mod(\x,4)}
  \ifnum\myresult=0\relax%
  \def\mycolor{blue}
  \fi%
  \ifnum\myresult=1\relax%
  \def\mycolor{green}
  \fi%
  \ifnum\myresult=2\relax%
  \def\mycolor{orange}
  \fi%
  \ifnum\myresult=3\relax%
  \def\mycolor{purple}
  \fi%
     \filldraw[draw=black,fill=\mycolor] (6.5, 0 - 5 + \x * 1 ) rectangle(9.5, 1 - 5 + \x * 1);
  
  \draw (16, 8.5) node {\Large CACHE};
  \foreach \x in {0,...,3}
  \pgfmathtruncatemacro\myresult{mod(\x,4)}
  \ifnum\myresult=0\relax%
  \def\mycolor{blue}
  \fi%
  \ifnum\myresult=1\relax%
  \def\mycolor{green}
  \fi%
  \ifnum\myresult=2\relax%
  \def\mycolor{orange}
  \fi%
  \ifnum\myresult=3\relax%
  \def\mycolor{purple}
  \fi%
     \filldraw[draw=black,fill=\mycolor] (13, 4 + \x * 1 ) rectangle(19, 5 + \x * 1);
\draw (16,4) -- (16,8);  
     \foreach \x in {4} \draw[->,>=latex] (9.5, - 5.5 + \x * 1) to[out=0,in=0] (19, 3.5 + \x * 1);
  \foreach \x in {1,...,4} \draw[->,>=latex] (9.5,  3.5 - \x * 1) to[out=30,in=210] (13, 8.5 - \x * 1);

  \foreach \x in {0,...,2} \draw[->,>=latex] (1.5,\x * 1 + 5.5) to[out=-35,in=180] (6.5,-0.5 + \x * 1);
     
  \foreach \x in {0,1} \draw[->,>=latex] (1.5,0.5 +\x * 1) to[out=0,in=180] (6.5, -1.5 +\x * 1 * 4);

\end{tikzpicture}
\label{fig:cache-colouring}
}
\subfloat[WCET Control Flow Graph]{
  \begin{tikzpicture}[thick,scale=0.20, every node/.style={transform shape}]

  \tikzstyle{noeud}=[minimum width=5cm,minimum height=3cm,rectangle,rounded corners=10pt,draw,text=black, text width= 2cm, text centered,font=\bfseries, ]

  \tikzstyle{noeud2}=[minimum width=5cm,minimum height=3cm,rectangle,rounded corners=10pt,draw,text=black,fill=lightgray, text width= 2cm, text centered,font=\bfseries, ]
  
  \node[noeud] (B1) at (0,0) {\Large BLOCK 1};
  \node[noeud] (B2) at (8,2) {\Large BLOCK 1};
  \node[noeud] (B3) at (8,-2) {\Large BLOCK 0};
  \node[noeud] (B4) at (16,0) {\Large BLOCK 1};

  \node[noeud] (B11) at (0,-8) {\Large BLOCK 1};
  \node[noeud] (B12) at (8,-6) {\Large BLOCK 1};
  \node[noeud2] (B13) at (8,-10) {\Large BLOCK 0};
  \node[noeud] (B14) at (16,-8) {\Large BLOCK 1};

  \draw[->,>=latex] (B1) edge[out=0, in=180] (B2)
  edge[out=0, in=180] (B3);

  \draw[<-,>=latex] (B4) edge[out=180, in=0] (B2)
  edge[out=180, in=0] (B3);

  \draw[->,>=latex] (B11) edge[out=0, in=180] (B12)
  edge[out=0, in=180] (B13);

  \draw[<-,>=latex] (B14) edge[out=180, in=0] (B12)
  edge[out=180, in=0] (B13);
  
\end{tikzpicture}
\label{fig:wcet-cfg}
}
\end{center}
\vspace{-1em}
\caption{Cache coloring and its impact on WCET}
\vspace{-1em}
\end{figure}
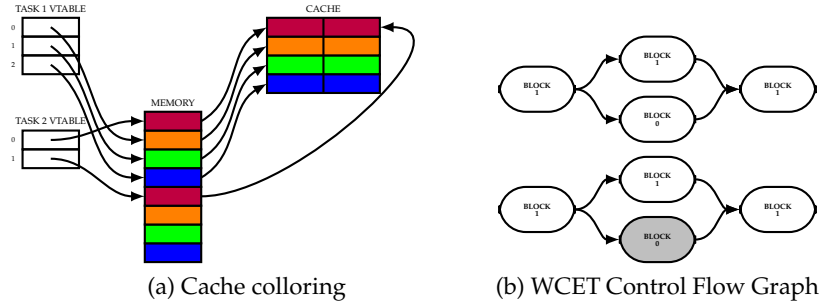

Main memory size is a multiple of cache memory size which is a
multiple of a page size. Therefore, when considering cache coloring
at page level, the same page is always mapped to the same cache page(partition of the cache memory of a page size).

Figure \ref{fig:cache-colouring} represents an example cache
coloring technique in a set-associative cache: all pages in main memory with
the same color share the same cache page (red color). Thus, the color
of each page in main memory can be computed as
$ \kappa_{i,j} = index(P_{i,j})mod\frac{S^{instructions}
}{N^{way}}$. $\kappa_{i,j}$ depends on the index of the page in
main memory, we can use the virtual table of the task to color
instructions pages. The configurations of task pages has an impact on
the typical execution of the task, thus also on the WCET.

\subsection{Recall on WCET analysis}
\label{sec:recall-wcet-analysis}

To compute the WCET, we can use measurements or static
analysis. Measurements give an optimistic estimation of the WCET
because not all inputs and internal states can be tested. The static
analysis gives a safer over-estimation of the WCET value. Our static
analysis method builds a control flow graph (CFG) of the task and
runs various analyses on it (including cache behavior prediction) to
compute an estimation of the WCET.  The task's pages allocation has
an impact on its WCET: in Figure \ref{fig:wcet-cfg} we show an example
of two CFGs of the same task with different page configurations. The
node's color represents the color of the page which contains the
block, and the edges the possible paths that the execution may
take. On the top CFG, Block 1 and Block 0 are not in the same page but
use the same area in the cache memory, if the wcet path uses Block 0, then
Block 1 will be evicted in cache memory. In the bottom CFG, Block 1
and Block 0 do not use the same cache memory area because the colors
of their pages are not the same, thus there will be no eviction and
lower execution time.


    \section{WCET-aware Coloring Heuristics}
\label{sec:algorithm}

Our goal is to allocate the virtual memory pages of a set of real-time
tasks to the physical memory pages so to minimize the inter-task
interference on the cache. In this paper, we try to completely
remove the interference by partitioning the cache.

We divide the problem into two steps: 1) 
  At the \emph{macro-level}, we assign a certain number of colors
  to each task so that the total number of colors is less than or
  equal to the number of available colors in the cache;
  2) At the \emph{micro-level}, for each task separately, with a given number
  of available colors, we compute the best WCET for that task.

We start by proposing a method for solving the \emph{micro-level}.

\subsection{Pages Coloring for a given partition size}
\label{sec:pages-coloring-given}

Consider that for each possible color combination, it is necessary to
perform a WCET analysis. Since that can be very time-consuming, we
rule out the complete exploration of all possible combination, and we
use a heuristic instead. An overestimation of the number of solutions
is given by $(P_i)^{P_i}$ which is exponential.

We consider 2 heuristic algorithms.
The first algorithm 
  assigns the same number of pages (approximately) to each
  color. In particular, if task $\tau_i$ is assigned $j$
  colors and it has $P_i$ pages, then the same color is assigned to
  $\lfloor P_i / j \rfloor$ pages. We use a simple modulo: the
  first $\lfloor P_i / j \rfloor $ pages are assigned to the
  first color, etc.
  
The second algorithm 
classifies pages according to their \emph{importance} in the
program. Therefore, we assign each page a \emph{score} that depends on
how many times the page is accessed by the program in the
Control-Flow-Graph. The score of a page is computed as the sum of the
scores of the instructions in the page, and the score of instruction
$\psi$ is computed as,
$score(\psi) =  10^{l(\psi)}$.
  Where $l(\psi)$ is the nesting level of loops where the instruction
  is found: if $\psi$ is not contained in any loop, then
  $l(\psi) = 0$; if $\psi$ is contained in a loop of first level, then
  $l(\psi) = 1$; etc.
  The pages' scores are computed using the OTAWA analysis tool
  \cite{ballabriga2010otawa}. Then the pages are ordered by decreasing
  score: if the task $\tau_i$ is assigned $j$ colors, the first $j-1$
  pages in decreasing score order are assigned a different color,
  while all other pages are assigned the last remaining color.

Once each page has been assigned a color according to one of the two
heuristics above, we launch the OTAWA WCET analysis tool to obtain the
corresponding WCET for the task.

We do this for different values of $j$ in the interval
$j = [1, S_i^{max}\}]$, where
\begin{equation}
  S_i^{max} = min\left\{P_i,S^{instructions} - (N-1)\right\}
  \label{equ:mx_cms_per_task}
\end{equation}
\noindent and for each value we compute the corresponding WCET
$C_i(j)$. These values are used by the ILP solver described in
the next section.

\subsection{Partitioning cache memory according to the need of the tasks}

The distribution of cache memory space can be represented as a
Multiple Choice Knapsack Problem(MCKP). In this problem we have a
knapsack of limited size and a set of objects of different
categories. The problem consists in selecting one and only one object
of each category to put in the knapsack.

In our case, the size of the knapsack represents the schedulability
constraints; the objective function is the number of colors used
(that we want to minimize); the object types are the different tasks,
and an object is a configuration of colors for a given task, with the
corresponding WCET.

We encode the problem above as an Integer Linear Programming (ILP)
problem, and we use CPLEX as a solver. We use the following variables
and constraints:

\begin{itemize}
\item We define variable $\chi_{i,j} \in \{0, 1\}$ to denote the fact
  that task $\tau_i$ has been assigned $j$ colors.  Each tasks must
  be assigned at least one configuration selected:
 $
      \sum_{j = 1}^{S_i^{max}} \chi_{i,j} = 1
 $

\item The worst case execution time of a task can be expressed as:
  $
    C_i = \sum_{j=1}^{S_i^{max}}(C_{i}(j)\cdot\chi_{i,j}).
  $
  where $C_i(j)$ is computed in the micro-level problem. 

\item We want to minimize the total number of colors used:
  $
    min~\sum_{i=1}^{N}\sum_{j = 1}^{S^{max}}  \left( j \cdot \chi_{i,j} \right)
  $
  If the value of the objective function for the optimal solution is
  greater than $\frac{S^{instructions}}{N^{way}}$, then the problem has no feasible
  solution, and we must resort to other methods for computing the
  interference (for example by using the CRPD analysis \cite{altmeyer2012improved}).

\item To impose the schedulability of the system, we use the DBF analysis
  for EDF, first proposed by Baruah \cite{baruah1990preemptively}. We
  first represent the utilization
  constraint 
  $
    \sum_{i=1}^{N} \frac{C_i}{T_i} \leq 1
    $
  Then we add all inequalities to check that all deadlines are
  respected. Let
  $dset(\tau_i) = \{\forall i=1, \ldots, N, \forall k > 0 | k T_i + D_i
  \leq DIT\}$. The first definitive IDLE time (DIT)
  \cite{lipari2013average}, is an instant at which all tasks must
  complete, and it does not depend on the WCETs of the tasks.
  Then, we add the following inequalities:
  $
    \forall t \in dset(\tau) : \quad \sum_{i=1}^{N}\left ( \left\lfloor \frac{t - D_i}{T_i} \right \rfloor + 1 \right) \cdot C_i\leq t
  $
\ifdefined\VLONGUE\else\vspace{-1em}\fi
\end{itemize}

    
    \section{Result}
\label{sec:result}

The analysis takes into account a system with a 32 KB set associative
memory cache of 2 ways with 512 rows. We consider a page size of 1 KB (this value is defined as a constant in OTAWA and involved timely tool modification if we want to change its value),
thus, there are 16 colors available.  We test each utilization in the
range $[0.30; 1.70]$ (we assume a step of $0.01$), with $1000$
variation of periods and deadlines of the 8 tasks in
Table~\ref{fig:array_tasks}, taken from well-known standard benchmarks
in the literature
\cite{falk2016taclebench,Gustafsson:WCET2010:Benchmarks}.

First, our method performs a static analysis of each task which gives
us a list of $WCET$ according to their number of available colors,
the worst of them is selected to compute the periods and deadline with
uunifast algorithm ($T_i = WCET_i(worst) / U_i$). To represents
constrained deadlines we assign for each task, a deadline in the
range of
$[WCET_i(worst) + (T_i - WCET_i(worst)) \cdot 0.75) \cdot T_i,
T_i]$. In the following figures, the line labeled as \emph{infinite
  cache} represents the percentage of schedulable tasks set that we can
schedule if we have a cache of unbounded size
\ifdefined\VLONGUE\else
with the WCET list from random coloring\fi.

The \emph{random} line represents the percentage of task schedulable
with a random distribution of the cache space between tasks
\ifdefined\VLONGUE\else and a random coloring of their pages\fi. Our
method (described in the previous section) is represented with the
\ifdefined\VLONGUE line \else lines \fi labeled \emph{ILP}. The x-axis
represents the utilization of the worst distribution with random
coloring.

\ifdefined\VLONGUE
\begin{figure}
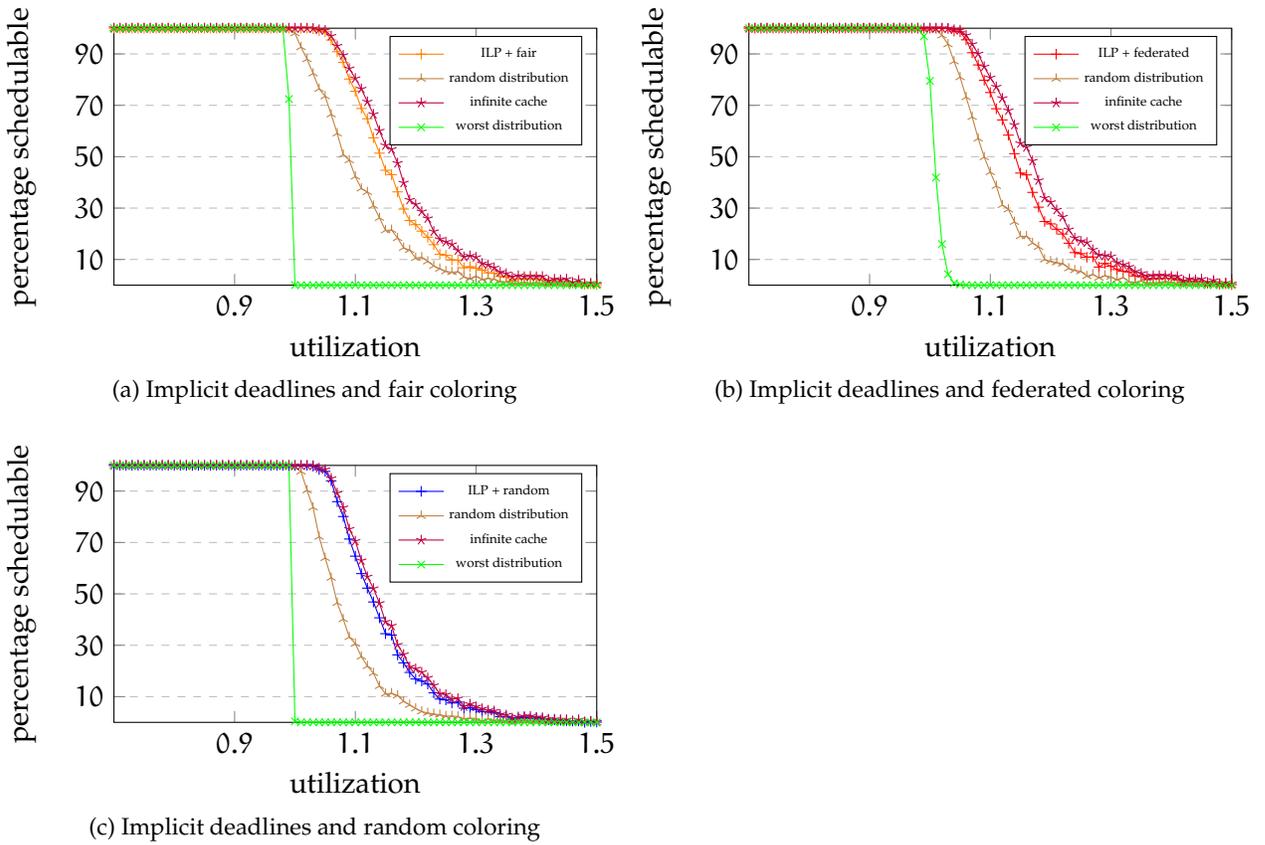

  \subfloat[Implicit deadlines and fair coloring]{\input{result_fair_i.tex}}
  \subfloat[Implicit deadlines and federated coloring]{\input{result_fed_i.tex}}

  \subfloat[Implicit deadlines and random coloring]{\input{result_rand_i.tex}}
  \caption{percentage of schedulable task set with implicit deadlines}
\end{figure}
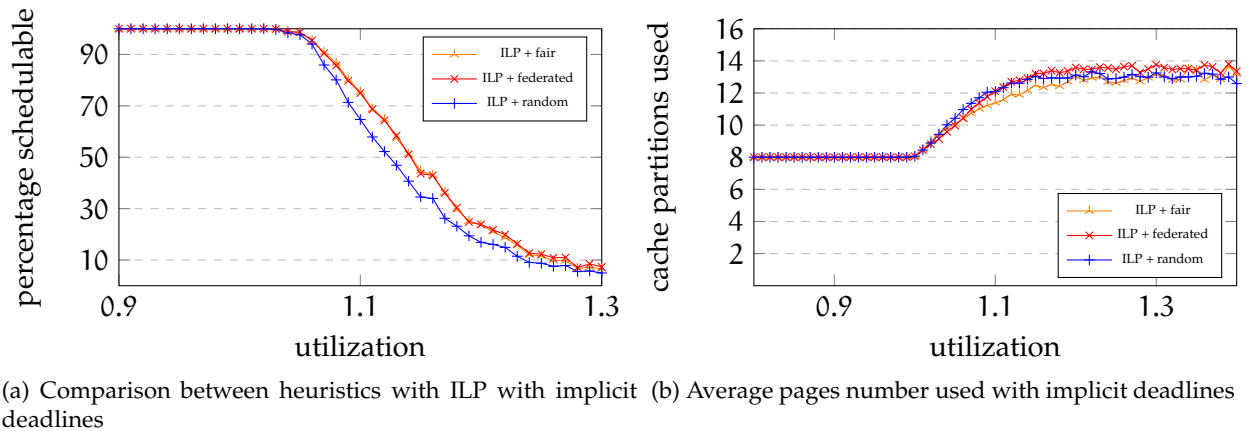
\begin{figure}
  \subfloat[Comparison between heuristics with ILP with implicit deadlines]{\begin{tikzpicture} 
\begin{axis}[legend style={font=\fontsize{5}{5}\selectfont},height=5cm,width=8cm,title={}, xlabel={utilization}, ylabel={percentage schedulable}, xmin=0.900000, xmax=1.300000, ymin=0.000000, ymax=100.000000,xtick={0.500000,0.700000,0.900000,1.100000,1.300000,1.500000},ytick={10.000000,30.000000,50.000000,70.000000,90.000000},legend pos=north east, ymajorgrids=true, grid style=dashed,]

\addplot[color=orange,mark=Mercedes star,] 
coordinates {(0.250000,100.000000)(0.260000,100.000000)(0.270000,100.000000)(0.280000,100.000000)(0.290000,100.000000)(0.300000,100.000000)(0.310000,100.000000)(0.320000,100.000000)(0.330000,100.000000)(0.340000,100.000000)(0.350000,100.000000)(0.360000,100.000000)(0.370000,100.000000)(0.380000,100.000000)(0.390000,100.000000)(0.400000,100.000000)(0.410000,100.000000)(0.420000,100.000000)(0.430000,100.000000)(0.440000,100.000000)(0.450000,100.000000)(0.460000,100.000000)(0.470000,100.000000)(0.480000,100.000000)(0.490000,100.000000)(0.500000,100.000000)(0.510000,100.000000)(0.520000,100.000000)(0.530000,100.000000)(0.540000,100.000000)(0.550000,100.000000)(0.560000,100.000000)(0.570000,100.000000)(0.580000,100.000000)(0.590000,100.000000)(0.600000,100.000000)(0.610000,100.000000)(0.620000,100.000000)(0.630000,100.000000)(0.640000,100.000000)(0.650000,100.000000)(0.660000,100.000000)(0.670000,100.000000)(0.680000,100.000000)(0.690000,100.000000)(0.700000,100.000000)(0.710000,100.000000)(0.720000,100.000000)(0.730000,100.000000)(0.740000,100.000000)(0.750000,100.000000)(0.760000,100.000000)(0.770000,100.000000)(0.780000,100.000000)(0.790000,100.000000)(0.800000,100.000000)(0.810000,100.000000)(0.820000,100.000000)(0.830000,100.000000)(0.840000,100.000000)(0.850000,100.000000)(0.860000,100.000000)(0.870000,100.000000)(0.880000,100.000000)(0.890000,100.000000)(0.900000,100.000000)(0.910000,100.000000)(0.920000,100.000000)(0.930000,100.000000)(0.940000,100.000000)(0.950000,100.000000)(0.960000,100.000000)(0.970000,100.000000)(0.980000,100.000000)(0.990000,100.000000)(1.000000,100.000000)(1.010000,100.000000)(1.020000,100.000000)(1.030000,99.807500)(1.040000,99.048500)(1.050000,98.609800)(1.060000,95.404400)(1.070000,90.856000)(1.080000,86.679500)(1.090000,80.221800)(1.100000,75.457200)(1.110000,68.756100)(1.120000,64.759300)(1.130000,57.363500)(1.140000,51.372500)(1.150000,44.715400)(1.160000,43.006700)(1.170000,36.302700)(1.180000,29.615700)(1.190000,25.164900)(1.200000,23.562700)(1.210000,21.028500)(1.220000,18.653800)(1.230000,15.764900)(1.240000,12.022900)(1.250000,11.617100)(1.260000,9.760590)(1.270000,9.746950)(1.280000,6.763290)(1.290000,7.199210)(1.300000,6.571430)(1.310000,5.882350)(1.320000,4.779760)(1.330000,4.635110)(1.340000,3.384910)(1.350000,3.154880)(1.360000,1.556780)(1.370000,2.417790)(1.380000,2.566450)(1.390000,2.543140)(1.400000,2.529180)(1.410000,1.823420)(1.420000,1.240460)(1.430000,1.488370)(1.440000,1.323830)(1.450000,1.224110)(1.460000,0.762631)(1.470000,1.120450)(1.480000,0.282486)(1.490000,0.487805)(1.500000,0.196850)(1.510000,0.093284)(1.520000,0.789733)(1.530000,0.559701)(1.540000,0.000000)(1.550000,0.101317)(1.560000,0.000000)(1.570000,0.000000)(1.580000,0.000000)(1.590000,0.000000)(1.600000,0.000000)(1.610000,0.000000)(1.620000,0.000000)(1.630000,0.000000)(1.640000,0.000000)(1.650000,0.000000)(1.660000,0.000000)}; 
\addlegendentry{ILP + fair}

\addplot[color=red,mark=x,] 
coordinates {(0.250000,100.000000)(0.260000,100.000000)(0.270000,100.000000)(0.280000,100.000000)(0.290000,100.000000)(0.300000,100.000000)(0.310000,100.000000)(0.320000,100.000000)(0.330000,100.000000)(0.340000,100.000000)(0.350000,100.000000)(0.360000,100.000000)(0.370000,100.000000)(0.380000,100.000000)(0.390000,100.000000)(0.400000,100.000000)(0.410000,100.000000)(0.420000,100.000000)(0.430000,100.000000)(0.440000,100.000000)(0.450000,100.000000)(0.460000,100.000000)(0.470000,100.000000)(0.480000,100.000000)(0.490000,100.000000)(0.500000,100.000000)(0.510000,100.000000)(0.520000,100.000000)(0.530000,100.000000)(0.540000,100.000000)(0.550000,100.000000)(0.560000,100.000000)(0.570000,100.000000)(0.580000,100.000000)(0.590000,100.000000)(0.600000,100.000000)(0.610000,100.000000)(0.620000,100.000000)(0.630000,100.000000)(0.640000,100.000000)(0.650000,100.000000)(0.660000,100.000000)(0.670000,100.000000)(0.680000,100.000000)(0.690000,100.000000)(0.700000,100.000000)(0.710000,100.000000)(0.720000,100.000000)(0.730000,100.000000)(0.740000,100.000000)(0.750000,100.000000)(0.760000,100.000000)(0.770000,100.000000)(0.780000,100.000000)(0.790000,100.000000)(0.800000,100.000000)(0.810000,100.000000)(0.820000,100.000000)(0.830000,100.000000)(0.840000,100.000000)(0.850000,100.000000)(0.860000,100.000000)(0.870000,100.000000)(0.880000,100.000000)(0.890000,100.000000)(0.900000,100.000000)(0.910000,100.000000)(0.920000,100.000000)(0.930000,100.000000)(0.940000,100.000000)(0.950000,100.000000)(0.960000,100.000000)(0.970000,100.000000)(0.980000,100.000000)(0.990000,100.000000)(1.000000,100.000000)(1.010000,100.000000)(1.020000,100.000000)(1.030000,99.903800)(1.040000,98.858200)(1.050000,98.609800)(1.060000,95.496300)(1.070000,90.466900)(1.080000,85.714300)(1.090000,79.759700)(1.100000,74.975900)(1.110000,68.658200)(1.120000,64.305200)(1.130000,58.393400)(1.140000,51.176500)(1.150000,43.631400)(1.160000,42.911500)(1.170000,36.111100)(1.180000,30.365500)(1.190000,24.787900)(1.200000,23.845400)(1.210000,21.763100)(1.220000,19.903800)(1.230000,16.324600)(1.240000,12.690800)(1.250000,12.267700)(1.260000,10.865600)(1.270000,10.965300)(1.280000,7.053140)(1.290000,8.481260)(1.300000,7.333330)(1.310000,6.078430)(1.320000,5.435800)(1.330000,5.128210)(1.340000,3.578340)(1.350000,3.346080)(1.360000,2.197800)(1.370000,2.611220)(1.380000,2.566450)(1.390000,2.724800)(1.400000,2.723740)(1.410000,2.111320)(1.420000,1.240460)(1.430000,1.488370)(1.440000,1.323830)(1.450000,1.412430)(1.460000,0.762631)(1.470000,1.307190)(1.480000,0.376648)(1.490000,0.585366)(1.500000,0.196850)(1.510000,0.093284)(1.520000,0.888450)(1.530000,0.559701)(1.540000,0.000000)(1.550000,0.101317)(1.560000,0.100806)(1.570000,0.000000)(1.580000,0.000000)(1.590000,0.000000)(1.600000,0.000000)(1.610000,0.000000)(1.620000,0.000000)(1.630000,0.000000)(1.640000,0.000000)(1.650000,0.000000)(1.660000,0.000000)}; 
\addlegendentry{ILP + federated}

  \addplot[color=blue,mark=+,] 
coordinates {(0.250000,100.000000)(0.260000,100.000000)(0.270000,100.000000)(0.280000,100.000000)(0.290000,100.000000)(0.300000,100.000000)(0.310000,100.000000)(0.320000,100.000000)(0.330000,100.000000)(0.340000,100.000000)(0.350000,100.000000)(0.360000,100.000000)(0.370000,100.000000)(0.380000,100.000000)(0.390000,100.000000)(0.400000,100.000000)(0.410000,100.000000)(0.420000,100.000000)(0.430000,100.000000)(0.440000,100.000000)(0.450000,100.000000)(0.460000,100.000000)(0.470000,100.000000)(0.480000,100.000000)(0.490000,100.000000)(0.500000,100.000000)(0.510000,100.000000)(0.520000,100.000000)(0.530000,100.000000)(0.540000,100.000000)(0.550000,100.000000)(0.560000,100.000000)(0.570000,100.000000)(0.580000,100.000000)(0.590000,100.000000)(0.600000,100.000000)(0.610000,100.000000)(0.620000,100.000000)(0.630000,100.000000)(0.640000,100.000000)(0.650000,100.000000)(0.660000,100.000000)(0.670000,100.000000)(0.680000,100.000000)(0.690000,100.000000)(0.700000,100.000000)(0.710000,100.000000)(0.720000,100.000000)(0.730000,100.000000)(0.740000,100.000000)(0.750000,100.000000)(0.760000,100.000000)(0.770000,100.000000)(0.780000,100.000000)(0.790000,100.000000)(0.800000,100.000000)(0.810000,100.000000)(0.820000,100.000000)(0.830000,100.000000)(0.840000,100.000000)(0.850000,100.000000)(0.860000,100.000000)(0.870000,100.000000)(0.880000,100.000000)(0.890000,100.000000)(0.900000,100.000000)(0.910000,100.000000)(0.920000,100.000000)(0.930000,100.000000)(0.940000,100.000000)(0.950000,100.000000)(0.960000,100.000000)(0.970000,100.000000)(0.980000,100.000000)(0.990000,100.000000)(1.000000,100.000000)(1.010000,100.000000)(1.020000,100.000000)(1.030000,99.711300)(1.040000,98.382500)(1.050000,97.683000)(1.060000,94.025700)(1.070000,85.894900)(1.080000,80.115800)(1.090000,71.349400)(1.100000,64.677600)(1.110000,57.884400)(1.120000,52.225200)(1.130000,46.858900)(1.140000,40.686300)(1.150000,34.507700)(1.160000,33.967600)(1.170000,26.245200)(1.180000,23.149000)(1.190000,19.415600)(1.200000,16.870900)(1.210000,16.069800)(1.220000,14.903800)(1.230000,11.473900)(1.240000,9.064890)(1.250000,8.736060)(1.260000,7.550640)(1.270000,7.872540)(1.280000,5.507250)(1.290000,5.719920)(1.300000,4.952380)(1.310000,4.117650)(1.320000,4.217430)(1.330000,3.846150)(1.340000,2.224370)(1.350000,2.294460)(1.360000,1.190480)(1.370000,1.644100)(1.380000,1.924840)(1.390000,1.453220)(1.400000,1.653700)(1.410000,1.343570)(1.420000,0.954198)(1.430000,0.930233)(1.440000,0.814664)(1.450000,0.659134)(1.460000,0.476644)(1.470000,0.560224)(1.480000,0.000000)(1.490000,0.292683)(1.500000,0.098425)(1.510000,0.093284)(1.520000,0.394867)(1.530000,0.373134)(1.540000,0.000000)(1.550000,0.000000)(1.560000,0.000000)(1.570000,0.000000)(1.580000,0.000000)(1.590000,0.000000)(1.600000,0.000000)(1.610000,0.000000)(1.620000,0.000000)(1.630000,0.000000)(1.640000,0.000000)(1.650000,0.000000)(1.660000,0.000000)}; 
\addlegendentry{ILP + random}

\end{axis}
\end{tikzpicture}

\label{fig:i_comparison}}
  \subfloat[Average pages number used with implicit deadlines]{\begin{tikzpicture} 
\begin{axis}[legend style={font=\fontsize{5}{5}\selectfont},height=5cm,width=8cm,title={}, xlabel={utilization}, ylabel={cache partitions used}, xmin=0.800000, xmax=1.400000, ymin=0.000000, ymax=16.000000,xtick={0.500000,0.700000,0.900000,1.100000,1.300000,1.500000},ytick={2.000000,4.000000,6.000000,8.000000,10.000000,12.000000,14.000000,16.000000},legend pos=south east, ymajorgrids=true, grid style=dashed,]

\addplot[color=orange,mark=Mercedes star] 
coordinates {(0.250000,8.000000)(0.260000,8.000000)(0.270000,8.000000)(0.280000,8.000000)(0.290000,8.000000)(0.300000,8.000000)(0.310000,8.000000)(0.320000,8.000000)(0.330000,8.000000)(0.340000,8.000000)(0.350000,8.000000)(0.360000,8.000000)(0.370000,8.000000)(0.380000,8.000000)(0.390000,8.000000)(0.400000,8.000000)(0.410000,8.000000)(0.420000,8.000000)(0.430000,8.000000)(0.440000,8.000000)(0.450000,8.000000)(0.460000,8.000000)(0.470000,8.000000)(0.480000,8.000000)(0.490000,8.000000)(0.500000,8.000000)(0.510000,8.000000)(0.520000,8.000000)(0.530000,8.000000)(0.540000,8.000000)(0.550000,8.000000)(0.560000,8.000000)(0.570000,8.000000)(0.580000,8.000000)(0.590000,8.000000)(0.600000,8.000000)(0.610000,8.000000)(0.620000,8.000000)(0.630000,8.000000)(0.640000,8.000000)(0.650000,8.000000)(0.660000,8.000000)(0.670000,8.000000)(0.680000,8.000000)(0.690000,8.000000)(0.700000,8.000000)(0.710000,8.000000)(0.720000,8.000000)(0.730000,8.000000)(0.740000,8.000000)(0.750000,8.000000)(0.760000,8.000000)(0.770000,8.000000)(0.780000,8.000000)(0.790000,8.000000)(0.800000,8.000000)(0.810000,8.000000)(0.820000,8.000000)(0.830000,8.000000)(0.840000,8.000000)(0.850000,8.000000)(0.860000,8.000000)(0.870000,8.000000)(0.880000,8.000000)(0.890000,8.000000)(0.900000,8.000000)(0.910000,8.000000)(0.920000,8.000000)(0.930000,8.000000)(0.940000,8.000000)(0.950000,8.000000)(0.960000,8.000000)(0.970000,8.000000)(0.980000,8.000000)(0.990000,8.000000)(1.000000,8.065780)(1.010000,8.484210)(1.020000,8.993350)(1.030000,9.422370)(1.040000,9.819400)(1.050000,10.084600)(1.060000,10.385400)(1.070000,10.721600)(1.080000,11.016700)(1.090000,11.231600)(1.100000,11.399200)(1.110000,11.582600)(1.120000,11.911600)(1.130000,11.892300)(1.140000,12.158400)(1.150000,12.486900)(1.160000,12.351800)(1.170000,12.559400)(1.180000,12.427200)(1.190000,12.704100)(1.200000,13.040000)(1.210000,12.799100)(1.220000,12.943300)(1.230000,13.011800)(1.240000,12.698400)(1.250000,12.624000)(1.260000,12.792500)(1.270000,12.942300)(1.280000,12.728600)(1.290000,12.958900)(1.300000,13.173900)(1.310000,13.116700)(1.320000,12.803900)(1.330000,12.851100)(1.340000,13.485700)(1.350000,13.424200)(1.360000,12.823500)(1.370000,13.120000)(1.380000,12.857100)(1.390000,13.642900)(1.400000,13.192300)(1.410000,13.631600)(1.420000,13.230800)(1.430000,13.687500)(1.440000,13.692300)(1.450000,14.615400)(1.460000,13.500000)(1.470000,13.833300)(1.480000,15.333300)(1.490000,14.400000)(1.500000,14.000000)(1.510000,13.000000)(1.520000,14.250000)(1.530000,14.333300)(1.540000,0.000000)(1.550000,16.000000)(1.560000,0.000000)(1.570000,0.000000)(1.580000,0.000000)(1.590000,0.000000)(1.600000,0.000000)(1.610000,0.000000)(1.620000,0.000000)(1.630000,0.000000)(1.640000,0.000000)(1.650000,0.000000)(1.660000,0.000000)}; 
\addlegendentry{ILP + fair}

\addplot[color=red,mark=x,] 
coordinates {(0.250000,8.000000)(0.260000,8.000000)(0.270000,8.000000)(0.280000,8.000000)(0.290000,8.000000)(0.300000,8.000000)(0.310000,8.000000)(0.320000,8.000000)(0.330000,8.000000)(0.340000,8.000000)(0.350000,8.000000)(0.360000,8.000000)(0.370000,8.000000)(0.380000,8.000000)(0.390000,8.000000)(0.400000,8.000000)(0.410000,8.000000)(0.420000,8.000000)(0.430000,8.000000)(0.440000,8.000000)(0.450000,8.000000)(0.460000,8.000000)(0.470000,8.000000)(0.480000,8.000000)(0.490000,8.000000)(0.500000,8.000000)(0.510000,8.000000)(0.520000,8.000000)(0.530000,8.000000)(0.540000,8.000000)(0.550000,8.000000)(0.560000,8.000000)(0.570000,8.000000)(0.580000,8.000000)(0.590000,8.000000)(0.600000,8.000000)(0.610000,8.000000)(0.620000,8.000000)(0.630000,8.000000)(0.640000,8.000000)(0.650000,8.000000)(0.660000,8.000000)(0.670000,8.000000)(0.680000,8.000000)(0.690000,8.000000)(0.700000,8.000000)(0.710000,8.000000)(0.720000,8.000000)(0.730000,8.000000)(0.740000,8.000000)(0.750000,8.000000)(0.760000,8.000000)(0.770000,8.000000)(0.780000,8.000000)(0.790000,8.000000)(0.800000,8.000000)(0.810000,8.000000)(0.820000,8.000000)(0.830000,8.000000)(0.840000,8.000000)(0.850000,8.000000)(0.860000,8.000000)(0.870000,8.000000)(0.880000,8.000000)(0.890000,8.000000)(0.900000,8.000000)(0.910000,8.000000)(0.920000,8.000000)(0.930000,8.000000)(0.940000,8.000000)(0.950000,8.000000)(0.960000,8.000000)(0.970000,8.000000)(0.980000,8.000000)(0.990000,8.000000)(1.000000,8.060060)(1.010000,8.433490)(1.020000,8.816540)(1.030000,9.151250)(1.040000,9.586140)(1.050000,9.969920)(1.060000,10.441800)(1.070000,10.974200)(1.080000,11.384000)(1.090000,11.764800)(1.100000,12.110400)(1.110000,12.365200)(1.120000,12.697700)(1.130000,12.793700)(1.140000,12.919500)(1.150000,13.209100)(1.160000,13.219500)(1.170000,13.395200)(1.180000,13.265400)(1.190000,13.357400)(1.200000,13.588900)(1.210000,13.481000)(1.220000,13.458900)(1.230000,13.605700)(1.240000,13.563900)(1.250000,13.477300)(1.260000,13.644100)(1.270000,13.700900)(1.280000,13.274000)(1.290000,13.500000)(1.300000,13.766200)(1.310000,13.580600)(1.320000,13.465500)(1.330000,13.519200)(1.340000,13.513500)(1.350000,13.428600)(1.360000,13.750000)(1.370000,13.629600)(1.380000,13.250000)(1.390000,13.800000)(1.400000,13.357100)(1.410000,14.045500)(1.420000,13.307700)(1.430000,13.187500)(1.440000,13.692300)(1.450000,14.266700)(1.460000,13.625000)(1.470000,13.571400)(1.480000,14.250000)(1.490000,14.166700)(1.500000,11.500000)(1.510000,13.000000)(1.520000,14.222200)(1.530000,14.166700)(1.540000,0.000000)(1.550000,15.000000)(1.560000,15.000000)(1.570000,0.000000)(1.580000,0.000000)(1.590000,0.000000)(1.600000,0.000000)(1.610000,0.000000)(1.620000,0.000000)(1.630000,0.000000)(1.640000,0.000000)(1.650000,0.000000)(1.660000,0.000000)}; 
\addlegendentry{ILP + federated}

\addplot[color=blue,mark=+,] 
coordinates {(0.250000,8.000000)(0.260000,8.000000)(0.270000,8.000000)(0.280000,8.000000)(0.290000,8.000000)(0.300000,8.000000)(0.310000,8.000000)(0.320000,8.000000)(0.330000,8.000000)(0.340000,8.000000)(0.350000,8.000000)(0.360000,8.000000)(0.370000,8.000000)(0.380000,8.000000)(0.390000,8.000000)(0.400000,8.000000)(0.410000,8.000000)(0.420000,8.000000)(0.430000,8.000000)(0.440000,8.000000)(0.450000,8.000000)(0.460000,8.000000)(0.470000,8.000000)(0.480000,8.000000)(0.490000,8.000000)(0.500000,8.000000)(0.510000,8.000000)(0.520000,8.000000)(0.530000,8.000000)(0.540000,8.000000)(0.550000,8.000000)(0.560000,8.000000)(0.570000,8.000000)(0.580000,8.000000)(0.590000,8.000000)(0.600000,8.000000)(0.610000,8.000000)(0.620000,8.000000)(0.630000,8.000000)(0.640000,8.000000)(0.650000,8.000000)(0.660000,8.000000)(0.670000,8.000000)(0.680000,8.000000)(0.690000,8.000000)(0.700000,8.000000)(0.710000,8.000000)(0.720000,8.000000)(0.730000,8.000000)(0.740000,8.000000)(0.750000,8.000000)(0.760000,8.000000)(0.770000,8.000000)(0.780000,8.000000)(0.790000,8.000000)(0.800000,8.000000)(0.810000,8.000000)(0.820000,8.000000)(0.830000,8.000000)(0.840000,8.000000)(0.850000,8.000000)(0.860000,8.000000)(0.870000,8.000000)(0.880000,8.000000)(0.890000,8.000000)(0.900000,8.000000)(0.910000,8.000000)(0.920000,8.000000)(0.930000,8.000000)(0.940000,8.000000)(0.950000,8.000000)(0.960000,8.000000)(0.970000,8.000000)(0.980000,8.000000)(0.990000,8.000000)(1.000000,8.064820)(1.010000,8.446890)(1.020000,8.887830)(1.030000,9.425680)(1.040000,10.018400)(1.050000,10.427900)(1.060000,10.966800)(1.070000,11.356700)(1.080000,11.703600)(1.090000,12.046600)(1.100000,12.102700)(1.110000,12.329900)(1.120000,12.617400)(1.130000,12.604400)(1.140000,12.848200)(1.150000,13.036600)(1.160000,12.899200)(1.170000,12.941600)(1.180000,12.931200)(1.190000,12.932000)(1.200000,13.117300)(1.210000,13.011400)(1.220000,13.316100)(1.230000,13.211400)(1.240000,12.873700)(1.250000,12.904300)(1.260000,12.987800)(1.270000,13.154800)(1.280000,13.017500)(1.290000,12.982800)(1.300000,13.250000)(1.310000,13.000000)(1.320000,12.888900)(1.330000,13.000000)(1.340000,13.000000)(1.350000,13.041700)(1.360000,13.230800)(1.370000,13.176500)(1.380000,12.857100)(1.390000,13.000000)(1.400000,12.588200)(1.410000,13.428600)(1.420000,13.500000)(1.430000,14.000000)(1.440000,13.250000)(1.450000,14.142900)(1.460000,14.200000)(1.470000,12.333300)(1.480000,0.000000)(1.490000,13.666700)(1.500000,14.000000)(1.510000,15.000000)(1.520000,14.250000)(1.530000,14.750000)(1.540000,0.000000)(1.550000,0.000000)(1.560000,0.000000)(1.570000,0.000000)(1.580000,0.000000)(1.590000,0.000000)(1.600000,0.000000)(1.610000,0.000000)(1.620000,0.000000)(1.630000,0.000000)(1.640000,0.000000)(1.650000,0.000000)(1.660000,0.000000)}; 
\addlegendentry{ILP + random}
\end{axis}
\end{tikzpicture}

\label{fig:i_pages}}
  \caption{performances of heuristics with implicit deadlines}
\end{figure}
\fi

\ifdefined\VLONGUE

\begin{figure} 
\subfloat[Constrained deadlines and fair coloring]{\input{result_fair_c.tex}}
\subfloat[Constrained deadlines and federated coloring]{\input{result_fed_c.tex}}

\subfloat[Constrained deadlines and random coloring]{\input{result_rand_c.tex}}
\subfloat[Tasks used in analysis]{
  \begin{tabular}[b]{ll}
    \hline
    Task & pages \\
    \hline
    compress (Mälardalen) & 4 \\
    fir (Mälardalen) & 2 \\
    ndes (Mälardalen) & 4 \\
    jfdctint (Mälardalen) & 3 \\
    edn (Mälardalen) & 4 \\
    crc (Mälardalen) & 2 \\
    g723\_enc (TACLeBench) & 8 \\
    petrinet (TACLeBench) & 8 \\
    \hline
\end{tabular}
\label{fig:array_tasks}
}
\caption{percentage of schedulable task set with constrained deadlines}
\end {figure}

\else

\begin{figure}%
  \input{heuristics.tex}%
  \vspace{-1em}%
  \caption{percentage of schedulable task set with constrained deadlines}%
  \label{fig:c_heuristics}%
  \vspace{-2em}%
\end{figure}%
\fi

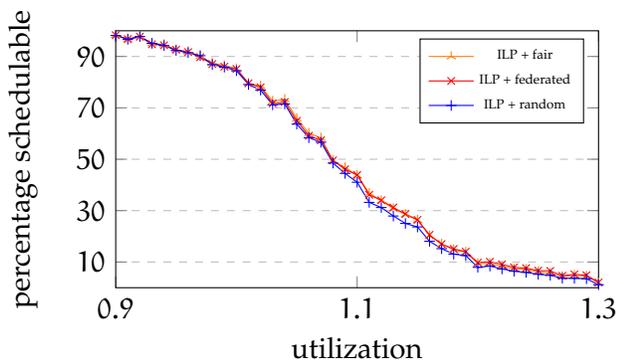
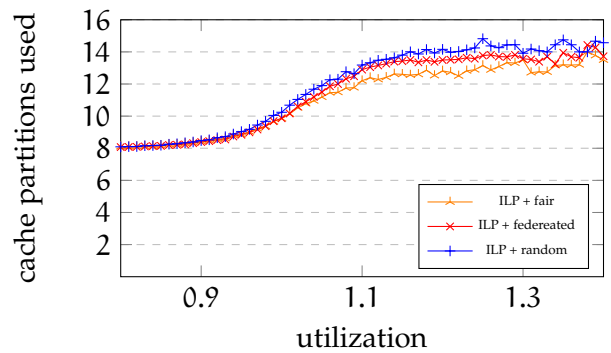
\begin{figure}
  \ifdefined\VLONGUE\subfloat[Comparison between heuristics with ILP
  with constrained deadlines]{\begin{tikzpicture} 
\begin{axis}[legend style={font=\fontsize{5}{5}\selectfont},height=5cm,width=8cm,title={}, xlabel={utilization}, ylabel={percentage schedulable}, xmin=0.900000, xmax=1.300000, ymin=0.000000, ymax=100.000000,xtick={0.500000,0.700000,0.900000,1.100000,1.300000,1.500000},ytick={10.000000,30.000000,50.000000,70.000000,90.000000},legend pos=north east, ymajorgrids=true, grid style=dashed,]

  \addplot[color=orange,mark=Mercedes star,] 
coordinates {(0.250000,100.000000)(0.260000,100.000000)(0.270000,100.000000)(0.280000,100.000000)(0.290000,100.000000)(0.300000,100.000000)(0.310000,100.000000)(0.320000,100.000000)(0.330000,100.000000)(0.340000,100.000000)(0.350000,100.000000)(0.360000,100.000000)(0.370000,100.000000)(0.380000,100.000000)(0.390000,100.000000)(0.400000,100.000000)(0.410000,100.000000)(0.420000,100.000000)(0.430000,100.000000)(0.440000,100.000000)(0.450000,100.000000)(0.460000,100.000000)(0.470000,100.000000)(0.480000,100.000000)(0.490000,100.000000)(0.500000,100.000000)(0.510000,100.000000)(0.520000,100.000000)(0.530000,100.000000)(0.540000,100.000000)(0.550000,100.000000)(0.560000,100.000000)(0.570000,100.000000)(0.580000,100.000000)(0.590000,100.000000)(0.600000,100.000000)(0.610000,100.000000)(0.620000,100.000000)(0.630000,100.000000)(0.640000,100.000000)(0.650000,100.000000)(0.660000,100.000000)(0.670000,100.000000)(0.680000,99.900800)(0.690000,99.909200)(0.700000,100.000000)(0.710000,100.000000)(0.720000,99.904800)(0.730000,100.000000)(0.740000,100.000000)(0.750000,99.807100)(0.760000,99.817400)(0.770000,99.900000)(0.780000,99.714600)(0.790000,99.718300)(0.800000,99.713700)(0.810000,99.705000)(0.820000,99.324300)(0.830000,99.514100)(0.840000,98.864700)(0.850000,98.293800)(0.860000,98.860400)(0.870000,98.390200)(0.880000,98.355900)(0.890000,97.513800)(0.900000,98.378900)(0.910000,96.857100)(0.920000,97.856500)(0.930000,95.115200)(0.940000,94.639600)(0.950000,92.627600)(0.960000,91.897400)(0.970000,90.105700)(0.980000,87.438200)(0.990000,86.489000)(1.000000,85.220700)(1.010000,79.562700)(1.020000,78.524100)(1.030000,72.598900)(1.040000,73.333300)(1.050000,65.799600)(1.060000,60.147600)(1.070000,58.294700)(1.080000,49.471700)(1.090000,46.748400)(1.100000,43.774300)(1.110000,36.736600)(1.120000,33.913000)(1.130000,30.747400)(1.140000,28.571400)(1.150000,26.393000)(1.160000,20.193200)(1.170000,16.979400)(1.180000,14.736800)(1.190000,13.832900)(1.200000,9.371980)(1.210000,9.399220)(1.220000,8.292680)(1.230000,7.348840)(1.240000,7.128910)(1.250000,6.168220)(1.260000,5.592420)(1.270000,4.347830)(1.280000,4.838710)(1.290000,4.491290)(1.300000,1.937980)(1.310000,2.986510)(1.320000,3.148150)(1.330000,1.467890)(1.340000,1.890360)(1.350000,1.447880)(1.360000,1.086960)(1.370000,0.963391)(1.380000,0.685602)(1.390000,0.843486)(1.400000,0.891266)(1.410000,0.675676)(1.420000,1.296110)(1.430000,0.584795)(1.440000,0.703518)(1.450000,0.374883)(1.460000,0.185874)(1.470000,0.000000)(1.480000,0.193424)(1.490000,0.093545)(1.500000,0.291262)(1.510000,0.098912)(1.520000,0.098717)(1.530000,0.000000)(1.540000,0.096246)(1.550000,0.000000)(1.560000,0.000000)(1.570000,0.000000)(1.580000,0.000000)(1.590000,0.000000)(1.600000,0.000000)(1.610000,0.000000)(1.620000,0.000000)(1.630000,0.000000)(1.640000,0.000000)(1.650000,0.000000)(1.660000,0.000000)(1.670000,0.000000)(19.110000,0.000000)}; 
\addlegendentry{ILP + fair}

\addplot[color=red,mark=x,] 
coordinates {(0.250000,100.000000)(0.260000,100.000000)(0.270000,100.000000)(0.280000,100.000000)(0.290000,100.000000)(0.300000,100.000000)(0.310000,100.000000)(0.320000,100.000000)(0.330000,100.000000)(0.340000,100.000000)(0.350000,100.000000)(0.360000,100.000000)(0.370000,100.000000)(0.380000,100.000000)(0.390000,100.000000)(0.400000,100.000000)(0.410000,100.000000)(0.420000,100.000000)(0.430000,100.000000)(0.440000,100.000000)(0.450000,100.000000)(0.460000,100.000000)(0.470000,100.000000)(0.480000,100.000000)(0.490000,100.000000)(0.500000,100.000000)(0.510000,100.000000)(0.520000,100.000000)(0.530000,100.000000)(0.540000,100.000000)(0.550000,100.000000)(0.560000,100.000000)(0.570000,100.000000)(0.580000,100.000000)(0.590000,100.000000)(0.600000,100.000000)(0.610000,100.000000)(0.620000,100.000000)(0.630000,100.000000)(0.640000,100.000000)(0.650000,100.000000)(0.660000,100.000000)(0.670000,100.000000)(0.680000,99.900800)(0.690000,99.818300)(0.700000,100.000000)(0.710000,100.000000)(0.720000,99.904800)(0.730000,100.000000)(0.740000,100.000000)(0.750000,99.807100)(0.760000,99.817400)(0.770000,99.800000)(0.780000,99.714600)(0.790000,99.718300)(0.800000,99.618300)(0.810000,99.705000)(0.820000,99.227800)(0.830000,99.514100)(0.840000,98.864700)(0.850000,98.293800)(0.860000,98.670500)(0.870000,98.390200)(0.880000,98.452600)(0.890000,97.513800)(0.900000,98.277600)(0.910000,96.761900)(0.920000,97.670100)(0.930000,94.746500)(0.940000,94.362300)(0.950000,92.627600)(0.960000,91.692300)(0.970000,89.721400)(0.980000,87.042500)(0.990000,86.029400)(1.000000,85.124800)(1.010000,79.562700)(1.020000,77.956500)(1.030000,71.751400)(1.040000,72.285700)(1.050000,64.932600)(1.060000,59.132800)(1.070000,57.367900)(1.080000,49.567700)(1.090000,46.088600)(1.100000,43.871600)(1.110000,36.164100)(1.120000,34.009700)(1.130000,31.220400)(1.140000,28.662400)(1.150000,26.393000)(1.160000,20.579700)(1.170000,16.979400)(1.180000,15.087700)(1.190000,14.025000)(1.200000,9.758450)(1.210000,10.077500)(1.220000,9.073170)(1.230000,7.906980)(1.240000,7.617190)(1.250000,6.635510)(1.260000,6.540280)(1.270000,4.637680)(1.280000,5.197130)(1.290000,4.857930)(1.300000,2.034880)(1.310000,3.468210)(1.320000,3.333330)(1.330000,1.834860)(1.340000,1.890360)(1.350000,1.737450)(1.360000,1.284580)(1.370000,1.059730)(1.380000,0.685602)(1.390000,1.030930)(1.400000,0.980392)(1.410000,0.675676)(1.420000,1.296110)(1.430000,0.584795)(1.440000,0.804020)(1.450000,0.374883)(1.460000,0.185874)(1.470000,0.000000)(1.480000,0.193424)(1.490000,0.093545)(1.500000,0.291262)(1.510000,0.197824)(1.520000,0.098717)(1.530000,0.000000)(1.540000,0.192493)(1.550000,0.000000)(1.560000,0.000000)(1.570000,0.000000)(1.580000,0.000000)(1.590000,0.000000)(1.600000,0.000000)(1.610000,0.000000)(1.620000,0.000000)(1.630000,0.000000)(1.640000,0.000000)(1.650000,0.000000)(1.660000,0.000000)(1.670000,0.000000)(19.110000,0.000000)}; 
\addlegendentry{ILP + federated}

\addplot[color=blue,mark=+,] 
coordinates {(0.250000,100.000000)(0.260000,100.000000)(0.270000,100.000000)(0.280000,100.000000)(0.290000,100.000000)(0.300000,100.000000)(0.310000,100.000000)(0.320000,100.000000)(0.330000,100.000000)(0.340000,100.000000)(0.350000,100.000000)(0.360000,100.000000)(0.370000,100.000000)(0.380000,100.000000)(0.390000,100.000000)(0.400000,100.000000)(0.410000,100.000000)(0.420000,100.000000)(0.430000,100.000000)(0.440000,100.000000)(0.450000,100.000000)(0.460000,100.000000)(0.470000,100.000000)(0.480000,100.000000)(0.490000,100.000000)(0.500000,100.000000)(0.510000,100.000000)(0.520000,100.000000)(0.530000,100.000000)(0.540000,100.000000)(0.550000,100.000000)(0.560000,100.000000)(0.570000,100.000000)(0.580000,100.000000)(0.590000,100.000000)(0.600000,100.000000)(0.610000,100.000000)(0.620000,100.000000)(0.630000,100.000000)(0.640000,100.000000)(0.650000,100.000000)(0.660000,100.000000)(0.670000,100.000000)(0.680000,99.900800)(0.690000,99.818300)(0.700000,100.000000)(0.710000,100.000000)(0.720000,99.904800)(0.730000,100.000000)(0.740000,100.000000)(0.750000,99.807100)(0.760000,99.908700)(0.770000,99.900000)(0.780000,99.714600)(0.790000,99.718300)(0.800000,99.618300)(0.810000,99.705000)(0.820000,99.324300)(0.830000,99.514100)(0.840000,98.959300)(0.850000,98.293800)(0.860000,98.670500)(0.870000,98.484800)(0.880000,98.452600)(0.890000,97.513800)(0.900000,98.176300)(0.910000,96.666700)(0.920000,97.763300)(0.930000,95.115200)(0.940000,94.269900)(0.950000,92.344000)(0.960000,91.487200)(0.970000,90.297800)(0.980000,86.844700)(0.990000,85.845600)(1.000000,84.453000)(1.010000,79.087500)(1.020000,76.915800)(1.030000,71.186400)(1.040000,71.523800)(1.050000,63.680200)(1.060000,58.394800)(1.070000,56.626500)(1.080000,48.511000)(1.090000,44.486300)(1.100000,41.050600)(1.110000,33.206100)(1.120000,31.207700)(1.130000,27.909200)(1.140000,25.022700)(1.150000,23.655900)(1.160000,18.067600)(1.170000,15.197000)(1.180000,13.070200)(1.190000,12.488000)(1.200000,7.922710)(1.210000,8.430230)(1.220000,7.317070)(1.230000,6.418600)(1.240000,5.957030)(1.250000,5.233640)(1.260000,4.834120)(1.270000,3.671500)(1.280000,3.673840)(1.290000,3.483040)(1.300000,1.162790)(1.310000,2.697500)(1.320000,2.500000)(1.330000,1.376150)(1.340000,1.701320)(1.350000,1.158300)(1.360000,0.889328)(1.370000,0.674374)(1.380000,0.293830)(1.390000,0.562324)(1.400000,0.623886)(1.410000,0.579151)(1.420000,0.997009)(1.430000,0.292398)(1.440000,0.402010)(1.450000,0.374883)(1.460000,0.092937)(1.470000,0.000000)(1.480000,0.000000)(1.490000,0.093545)(1.500000,0.097087)(1.510000,0.098912)(1.520000,0.000000)(1.530000,0.000000)(1.540000,0.000000)(1.550000,0.000000)(1.560000,0.000000)(1.570000,0.000000)(1.580000,0.000000)(1.590000,0.000000)(1.600000,0.000000)(1.610000,0.000000)(1.620000,0.000000)(1.630000,0.000000)(1.640000,0.000000)(1.650000,0.000000)(1.660000,0.000000)(1.670000,0.000000)(19.110000,0.000000)}; 
\addlegendentry{ILP + random}

\end{axis}
\end{tikzpicture}

\label{fig:c_comparison}}\fi
  \subfloat[Average number of cache partitions used with constrained
  deadlines]{\begin{tikzpicture} 
\begin{axis}[legend style={font=\fontsize{5}{5}\selectfont},height=5cm,width=8cm,title={}, xlabel={utilization}, ylabel={cache partitions used}, xmin=0.800000, xmax=1.400000, ymin=0.000000, ymax=16.000000,xtick={0.500000,0.700000,0.900000,1.100000,1.300000,1.500000},ytick={2.000000,4.000000,6.000000,8.000000,10.000000,12.000000,14.000000,16.000000},legend pos=south east, ymajorgrids=true, grid style=dashed,]

\addplot[color=orange,mark=Mercedes star,] 
coordinates {(0.250000,8.000000)(0.260000,8.000000)(0.270000,8.000000)(0.280000,8.000000)(0.290000,8.000000)(0.300000,8.000000)(0.310000,8.000000)(0.320000,8.000000)(0.330000,8.000000)(0.340000,8.000000)(0.350000,8.000000)(0.360000,8.000000)(0.370000,8.000000)(0.380000,8.000000)(0.390000,8.000000)(0.400000,8.000000)(0.410000,8.000000)(0.420000,8.000000)(0.430000,8.000000)(0.440000,8.000000)(0.450000,8.000000)(0.460000,8.000000)(0.470000,8.000000)(0.480000,8.000000)(0.490000,8.000000)(0.500000,8.000000)(0.510000,8.000000)(0.520000,8.000000)(0.530000,8.000000)(0.540000,8.000000)(0.550000,8.000000)(0.560000,8.000000)(0.570000,8.000000)(0.580000,8.000000)(0.590000,8.000000)(0.600000,8.000000)(0.610000,8.000000)(0.620000,8.002860)(0.630000,8.000000)(0.640000,8.000000)(0.650000,8.003830)(0.660000,8.001910)(0.670000,8.005050)(0.680000,8.002980)(0.690000,8.011820)(0.700000,8.000940)(0.710000,8.012130)(0.720000,8.017160)(0.730000,8.034650)(0.740000,8.011260)(0.750000,8.020290)(0.760000,8.041170)(0.770000,8.023020)(0.780000,8.024810)(0.790000,8.051790)(0.800000,8.101440)(0.810000,8.097630)(0.820000,8.112730)(0.830000,8.132810)(0.840000,8.134930)(0.850000,8.162010)(0.860000,8.238230)(0.870000,8.238690)(0.880000,8.294000)(0.890000,8.342780)(0.900000,8.425330)(0.910000,8.476890)(0.920000,8.560000)(0.930000,8.608530)(0.940000,8.798830)(0.950000,8.858160)(0.960000,9.021210)(0.970000,9.224950)(0.980000,9.365380)(0.990000,9.656750)(1.000000,9.844590)(1.010000,10.131400)(1.020000,10.519300)(1.030000,10.779500)(1.040000,10.955800)(1.050000,11.185900)(1.060000,11.460100)(1.070000,11.486500)(1.080000,11.747600)(1.090000,11.774200)(1.100000,12.204400)(1.110000,12.394800)(1.120000,12.284900)(1.130000,12.415400)(1.140000,12.627400)(1.150000,12.603700)(1.160000,12.559800)(1.170000,12.602200)(1.180000,12.857100)(1.190000,12.527800)(1.200000,12.814400)(1.210000,12.732000)(1.220000,12.505900)(1.230000,12.822800)(1.240000,12.890400)(1.250000,13.151500)(1.260000,12.881400)(1.270000,13.088900)(1.280000,13.333300)(1.290000,13.306100)(1.300000,13.500000)(1.310000,12.709700)(1.320000,12.764700)(1.330000,12.750000)(1.340000,13.150000)(1.350000,13.200000)(1.360000,13.181800)(1.370000,13.200000)(1.380000,14.000000)(1.390000,13.777800)(1.400000,13.500000)(1.410000,13.000000)(1.420000,14.076900)(1.430000,14.666700)(1.440000,13.571400)(1.450000,13.750000)(1.460000,14.500000)(1.470000,0.000000)(1.480000,15.500000)(1.490000,14.000000)(1.500000,14.333300)(1.510000,15.000000)(1.520000,14.000000)(1.530000,0.000000)(1.540000,14.000000)(1.550000,0.000000)(1.560000,0.000000)(1.570000,0.000000)(1.580000,0.000000)(1.590000,0.000000)(1.600000,0.000000)(1.610000,0.000000)(1.620000,0.000000)(1.630000,0.000000)(1.640000,0.000000)(1.650000,0.000000)(1.660000,0.000000)(1.670000,0.000000)(19.110000,0.000000)}; 
\addlegendentry{ILP + fair}

\addplot[color=red,mark=x,] 
coordinates {(0.250000,8.000000)(0.260000,8.000000)(0.270000,8.000000)(0.280000,8.000000)(0.290000,8.000000)(0.300000,8.000000)(0.310000,8.000000)(0.320000,8.000000)(0.330000,8.000000)(0.340000,8.000000)(0.350000,8.000000)(0.360000,8.000000)(0.370000,8.000000)(0.380000,8.000000)(0.390000,8.000000)(0.400000,8.000000)(0.410000,8.000000)(0.420000,8.000000)(0.430000,8.000000)(0.440000,8.000000)(0.450000,8.000000)(0.460000,8.000000)(0.470000,8.000000)(0.480000,8.000000)(0.490000,8.000000)(0.500000,8.000000)(0.510000,8.000000)(0.520000,8.000000)(0.530000,8.000000)(0.540000,8.000000)(0.550000,8.000000)(0.560000,8.000000)(0.570000,8.000000)(0.580000,8.000000)(0.590000,8.000000)(0.600000,8.000000)(0.610000,8.000000)(0.620000,8.000950)(0.630000,8.000000)(0.640000,8.000000)(0.650000,8.000960)(0.660000,8.000950)(0.670000,8.006060)(0.680000,8.001990)(0.690000,8.004550)(0.700000,8.000940)(0.710000,8.009330)(0.720000,8.013350)(0.730000,8.029700)(0.740000,8.012200)(0.750000,8.014490)(0.760000,8.031110)(0.770000,8.012020)(0.780000,8.020990)(0.790000,8.048020)(0.800000,8.071840)(0.810000,8.078900)(0.820000,8.085600)(0.830000,8.111330)(0.840000,8.131100)(0.850000,8.146580)(0.860000,8.203080)(0.870000,8.215590)(0.880000,8.238700)(0.890000,8.317280)(0.900000,8.386600)(0.910000,8.443900)(0.920000,8.507630)(0.930000,8.544750)(0.940000,8.730660)(0.950000,8.838780)(0.960000,8.988810)(0.970000,9.165950)(0.980000,9.376140)(0.990000,9.685900)(1.000000,9.866970)(1.010000,10.164900)(1.020000,10.603200)(1.030000,10.893700)(1.040000,11.196300)(1.050000,11.500000)(1.060000,11.858000)(1.070000,12.000000)(1.080000,12.327500)(1.090000,12.482600)(1.100000,12.895800)(1.110000,13.050100)(1.120000,13.156200)(1.130000,13.266700)(1.140000,13.393700)(1.150000,13.425900)(1.160000,13.497700)(1.170000,13.348100)(1.180000,13.494200)(1.190000,13.383600)(1.200000,13.495000)(1.210000,13.509600)(1.220000,13.537600)(1.230000,13.635300)(1.240000,13.576900)(1.250000,13.774600)(1.260000,13.811600)(1.270000,13.708300)(1.280000,13.689700)(1.290000,13.811300)(1.300000,13.571400)(1.310000,13.527800)(1.320000,13.388900)(1.330000,13.700000)(1.340000,13.250000)(1.350000,13.944400)(1.360000,13.692300)(1.370000,13.636400)(1.380000,14.428600)(1.390000,14.272700)(1.400000,13.727300)(1.410000,13.285700)(1.420000,13.846200)(1.430000,14.500000)(1.440000,14.250000)(1.450000,13.250000)(1.460000,14.000000)(1.470000,0.000000)(1.480000,15.000000)(1.490000,14.000000)(1.500000,14.000000)(1.510000,15.000000)(1.520000,14.000000)(1.530000,0.000000)(1.540000,14.500000)(1.550000,0.000000)(1.560000,0.000000)(1.570000,0.000000)(1.580000,0.000000)(1.590000,0.000000)(1.600000,0.000000)(1.610000,0.000000)(1.620000,0.000000)(1.630000,0.000000)(1.640000,0.000000)(1.650000,0.000000)(1.660000,0.000000)(1.670000,0.000000)(19.110000,0.000000)}; 
\addlegendentry{ILP + federeated}

\addplot[color=blue,mark=+,] 
coordinates {(0.250000,8.000000)(0.260000,8.000000)(0.270000,8.000000)(0.280000,8.000000)(0.290000,8.000000)(0.300000,8.000000)(0.310000,8.000000)(0.320000,8.000000)(0.330000,8.000000)(0.340000,8.000000)(0.350000,8.000000)(0.360000,8.000000)(0.370000,8.000000)(0.380000,8.000000)(0.390000,8.000000)(0.400000,8.000000)(0.410000,8.000000)(0.420000,8.000000)(0.430000,8.000000)(0.440000,8.000000)(0.450000,8.000000)(0.460000,8.000000)(0.470000,8.000000)(0.480000,8.000000)(0.490000,8.000000)(0.500000,8.000000)(0.510000,8.000000)(0.520000,8.000000)(0.530000,8.000000)(0.540000,8.000000)(0.550000,8.000000)(0.560000,8.000000)(0.570000,8.000000)(0.580000,8.000000)(0.590000,8.000000)(0.600000,8.000000)(0.610000,8.000000)(0.620000,8.002860)(0.630000,8.000000)(0.640000,8.000000)(0.650000,8.002870)(0.660000,8.001910)(0.670000,8.005050)(0.680000,8.002980)(0.690000,8.006370)(0.700000,8.000940)(0.710000,8.013990)(0.720000,8.017160)(0.730000,8.033660)(0.740000,8.015010)(0.750000,8.020290)(0.760000,8.045700)(0.770000,8.025030)(0.780000,8.032440)(0.790000,8.056500)(0.800000,8.088120)(0.810000,8.099610)(0.820000,8.104960)(0.830000,8.157230)(0.840000,8.168260)(0.850000,8.186110)(0.860000,8.252170)(0.870000,8.279810)(0.880000,8.340860)(0.890000,8.389990)(0.900000,8.474720)(0.910000,8.525120)(0.920000,8.653960)(0.930000,8.720930)(0.940000,8.885290)(0.950000,9.034800)(0.960000,9.176010)(0.970000,9.444680)(0.980000,9.675400)(0.990000,10.043900)(1.000000,10.239800)(1.010000,10.681500)(1.020000,11.039400)(1.030000,11.350500)(1.040000,11.675100)(1.050000,11.900200)(1.060000,12.265400)(1.070000,12.301100)(1.080000,12.766300)(1.090000,12.641900)(1.100000,13.180100)(1.110000,13.324700)(1.120000,13.476800)(1.130000,13.539000)(1.140000,13.632700)(1.150000,13.797500)(1.160000,14.005300)(1.170000,13.870400)(1.180000,14.147700)(1.190000,13.930800)(1.200000,14.158500)(1.210000,13.977000)(1.220000,14.026700)(1.230000,14.130400)(1.240000,14.245900)(1.250000,14.821400)(1.260000,14.372500)(1.270000,14.263200)(1.280000,14.439000)(1.290000,14.447400)(1.300000,13.916700)(1.310000,14.178600)(1.320000,14.074100)(1.330000,14.000000)(1.340000,14.444400)(1.350000,14.750000)(1.360000,14.444400)(1.370000,14.000000)(1.380000,14.000000)(1.390000,14.666700)(1.400000,14.571400)(1.410000,14.333300)(1.420000,15.100000)(1.430000,15.333300)(1.440000,14.500000)(1.450000,15.250000)(1.460000,13.000000)(1.470000,0.000000)(1.480000,0.000000)(1.490000,16.000000)(1.500000,16.000000)(1.510000,16.000000)(1.520000,0.000000)(1.530000,0.000000)(1.540000,0.000000)(1.550000,0.000000)(1.560000,0.000000)(1.570000,0.000000)(1.580000,0.000000)(1.590000,0.000000)(1.600000,0.000000)(1.610000,0.000000)(1.620000,0.000000)(1.630000,0.000000)(1.640000,0.000000)(1.650000,0.000000)(1.660000,0.000000)(1.670000,0.000000)(19.110000,0.000000)}; 
\addlegendentry{ILP + random}

\end{axis}
\end{tikzpicture}

\label{fig:c_pages}} \ifdefined\VLONGUE\else \hfill
  \subfloat[Tasks used in analysis]{
    \begin{tabular}[b]{ll}
      \hline
      Task & pages \\
      \hline 
      compress (Mälardalen) & 4 \\
      fir (Mälardalen) & 2 \\
      ndes (Mälardalen) & 4 \\
      jfdctint (Mälardalen) & 3 \\
      edn (Mälardalen) & 4 \\
      crc (Mälardalen) & 2 \\
      g723\_enc (TACLeBench) & 8 \\
      petrinet (TACLeBench) & 8 \\
      \hline 
    \end{tabular}
    \label{fig:array_tasks}
    
}
\fi
\caption{performances of heuristics with constrained deadlines}
\ifdefined\VLONGUE\else\vspace{-2em}\fi
\end{figure}
For all heuristics,
\ifdefined\VLONGUE
Figures~\ref{fig:i_fair}, \ref{fig:i_fed}, \ref{fig:i_rand}, \ref{fig:c_fair}, \ref{fig:c_fed} and
\ref{fig:c_rand} show
\else
Figure~\ref{fig:c_heuristics} shows
\fi
that our method (ILP)
increases the amount of schedulable set (more than 20\% compared to
random distribution for high utilization), but the performances of our
heuristics are mitigated compared to the random coloring\ifdefined\VLONGUE
 (see Figure~\ref{fig:c_comparison})
\fi
.  On this
figure, we do not observe any significant difference between the
performance of fair coloring and federated coloring. However, if we
look at Figure~\ref{fig:c_pages} we can see that fair coloring uses
fewer pages than federated and random. So the best heuristics is fair
coloring compared to random and federated. This can be explained by
the fact that for a low number of colors $j$, federated coloring
isolates only the $j-1$ most important pages. If the score of the
$j$-th pages is also important, it will experience a significant number
of evictions in all the other pages with the lower scores.


    \section{Conclusion}
\ifdefined\VLONGUE\else\vspace{-0.5em}\fi
We proposed an approach based on ILP to partition the cache memory
according to the needs of each task for a preemptive system scheduled
with EDF. We also propose a heuristic based on our empirical results
to find pages layout for each task according to the number of its
given colors.  Our experimental results confirm an increase of high
utilization tasks set schedulable of $20\%$ compared to a random
partition of cache memory, however, the performances of our heuristics
to coloring task pages are mitigated. We will reduce the granularity
to have a method to partition at the granularity of the size of a
cache line and explore other heuristics in a future work.

    \bibliography{biblio} 

\end{document}